\documentclass[prd, aps, nobibnotes, nofootinbib,
eqsecnum,
preprint,
amsmath,
amssymb]{revtex4-2}

\usepackage{multirow}
\usepackage[utf8]{inputenc}
\usepackage[T1]{fontenc}
\usepackage{physics}
\usepackage{amsmath}
\usepackage{braket}
\usepackage{pgfplots}
\usepackage{subfig}
\usepackage{commath}
\usepackage{feynmp-auto}
\usepackage{bm}
\usepackage{hyperref}
\usepackage{slashed}
\usepackage{centernot}
\usepackage{bbold}
\usepackage{tensor}
\usepackage{amssymb}
\usepackage{mathrsfs}
\usepackage{mathtools}
\usepackage{tikz-cd}
\usetikzlibrary{positioning, shapes, shadows, arrows}
 \usepackage[compat=1.0.0]{tikz-feynman}
\usepackage{graphicx}
\usepackage{xspace}
\graphicspath{ {./p/} }

\pgfplotsset{compat=1.18} 

\begin{document}

\title{Parity nonconservation induced by spacetime geometry}

\author{Arnab Chakraborty}
\email{arnabchakraborty233@bose.res.in}
\author{Amitabha Lahiri}
\email{amitabha@bose.res.in}
\affiliation{Satyendra Nath Bose National Centre for Basic Sciences\\
	Block - JD, Sector - III, Salt Lake, Kolkata - 700106}

\begin{abstract}
	The interaction of fermion spin with spacetime can be non-universal, leading to a new interaction beyond the Standard Model, independent of gravitation.	
	Fermions generate spacetime torsion, which can be integrated out in favor of a four-fermion interaction in a torsion-free background. This is a current-current interaction which involves all fermions and generically has different coupling constants for different chiralities and species of fermions.  It does not vanish when curvature goes to zero, so accelerator experiments should be able to see its effect. We calculate the contribution of this geometrical interaction to parity nonconservation in $e^-e^-$ and $e^-D$ scattering and compare with known observations. This provides an estimate of an upper bound on the coupling constants, suggesting that the strength of the ``new physics'' can be as large as only one order of magnitude smaller than that of weak interactions. 
\end{abstract}

\maketitle

\newpage

\section{Introduction}
It is well known that fermions induce torsion in spacetime. This torsion is an independent non-dynamical field and is fully determined by the fermions in the theory. Thus it can be eliminated from the action. The result is ordinary torsion-free gravity, but with a four-fermion current-current interaction term. This is a kind of ``new physics'' which does not involve any new symmetry or new particles beyond the Standard Model. As we point out below, this interaction affects all species of fermions and has non-universal couplings which cannot be fixed from theoretical arguments. The interaction can also be different for left- and right-handed fermions, thus naturally breaking left-right symmetry. Thus, among the many processes which may be affected by this ``geometrical'' or `torsional'' interaction are those which violate parity. In this paper we will investigate the effect of the geometrical interaction on some parity-violating processes and try to estimate an upper bound on the scale of the couplings of this interaction.

Parity violation has been known in the context of weak interaction for a long time -- the original suggestion and observation~\cite{Lee:1956qn, Wu:1957my} 
came in the context of nuclear beta decay, even before there was an established theory of electroweak interactions. The Standard Model introduced weak 
neutral current, which allowed parity nonconservation in elastic scattering~\cite{Derman:1979zc}. {Reviews of polarized electron scattering and the corresponding parity-violating asymmetry can be found in ~\cite{Souder:2016zc,Kumar:2005lzv,Kumar:2011lzv,Kumar:2012lzv}.} Measurements of parity-violating asymmetry have been reported for elastic scattering of polarized electrons from unpolarized electrons~\cite{Anthony:2005zc, SLACE158:2003onx} an unpolarized target such as the proton~\cite{Spayde:2000zc,Androic:2013zc}, deuteron~\cite{Wang:2013zc, Wang:2014zc, Wang:2015zc}, or the nuclei C-12~\cite{Souder:1990zc}, Ca-48~\cite{Adhikari:2022zc} and Pb-208~\cite{Abrahamyan:2012zc,Adhikari:2021zc}.  
In contrast to the beta decay experiments, these show very tiny parity violating asymmetries, of the order of $10^{-6}$ or smaller. The geometrical 
four-fermion interaction that we consider is a current-current interaction which does not change the identities of the particles, so we can expect 
it to contribute to the same processes which are mediated by the neutral current. The scale of this interaction is unknown, but we can estimate an 
upper bound on it by calculating the parity asymmetry after including this interaction and comparing the result with observed values. 
In this paper, we will make an 
order of magnitude estimate of the scale of the geometrical interaction. 

The paper is organized as follows. In Sec.~\ref{origin} we briefly discuss the origin of the geometrical interaction and write the interaction term.
In Sec.~\ref{Moller} and Sec.~\ref{e-D} we calculate the parity asymmetry in M{\o}ller scattering and electron-deuteron scattering, respectively. 
In the last section, we calculate some bounds on the scale of the coupling constants of the geometrical interaction, based on the comparison of 
our results with experimental values.

\section{Geometrical four-fermion interaction}\label{origin}
The geometrical interaction is a direct consequence of the dynamics of fermions on curved spacetime, which we now briefly review. We will employ the Einstein-Cartan-Sciama-Kibble (ECSK) formalism, which is very convenient for describing fermions on a curved background~\cite{Cartan:1923zea, Cartan:1924yea, Gasperini:2013,Hehl:1974cn,Hammond:2002rm,Hehl:1976kj, Hehl:2007bn, Kibble:1961ba, Mielke:2017nwt, Sciama:1964wt, Poplawski:2009fb, Chakrabarty:2018ybk}. In this formalism,  Dirac's $\gamma$ matrices are defined on an ``internal'' flat space, isomorphic to the tangent space at each point. Then we have $\left[\gamma^a, \gamma^b\right]_{+} = 2\eta^{ab}$\,. At each point there are four orthonormal vector fields  $e^\mu_a$ called tetrads, and their inverses $e^a_\mu$\, called co-tetrads, satisfying
\begin{align}
	g_{\mu\nu}e^\mu_a e^\nu_b = \eta_{ab}\,, \quad \eta_{{ab}}e^a_\mu e^b_\nu = g_{\mu\nu}\,, \quad 
	\quad e^\mu_a e^a_\nu = \delta^\mu_\nu\,.
\end{align}
The spacetime indices will be denoted by lowercase Greek letters $\mu, \nu,\cdots$ and internal flat space indices by lowercase Latin letters $a, b, \cdots$\,. 
We will also define spacetime $\gamma$ matrices by $\gamma^\mu := e^\mu_a \gamma^a$ -- these satisfy $\left[\gamma_\mu\,, \gamma_\nu\right]_+ = 2g_{\mu\nu}$\,. We will sometimes use $\gamma^\mu$ if the combination $e^\mu_a \gamma^a$ appears in an equation. The co-tetrad, thought of as a 4$\times$4 matrix, has determinant equal to the square root of the matrix determinant, $|e| = \sqrt{|g|}$\,.  

The connection is assumed to annihilate the  spacetime $\gamma$ matrices, i.e. $\nabla_\mu \gamma_\nu = 0\,,$ so as to guarantee metric compatibility.
Then it must annihilate the tetrads, $\nabla_\mu e_\nu^a = 0\,.$ The connection has two parts, one for the spacetime and one for the internal space, so we can write
\begin{equation}
 \nabla_\mu e_\nu^a =\partial_\mu e^a_\nu + 	A_{\mu}{}^{a}{}_{b} e^b_\nu - \Gamma^\lambda{}_{\mu \nu}e_\lambda^a = 0\,.
	\label{tetrad-postulate}
\end{equation}
This equation is often referred to as the \textit{tetrad postulate}.  $A_{\mu}{}^{a}{}_{b}$ are components of the \textit{spin connection}, which appears in the covariant derivative of spinor fields when we write the Dirac operator,
\begin{equation}\label{Dirac-operator}
	D_\mu\psi = \partial_\mu\psi -\frac{i}{4} A_\mu{}^{ab} \sigma_{ab}\psi\,, \qquad \sigma_{ab} = \frac{i}{2}\left[\gamma_a\,, \gamma_b\right]_{-}\,.
\end{equation}
We will be using the signature $(+---)$ in this paper. 

The ``field strength'' of the spin connection is defined as 
\begin{equation}\label{F(A)}
	F_{\mu\nu}{}^{ab}(A) = \partial_\mu A_\nu{}^{ab} - \partial_\nu A_\mu{}^{ab} + A_{\mu}{}^{a}{}_{c} A_{\nu}{}^{cb} -  A_{\nu}{}^{a}{}_{c} A_{\mu}{}^{cb}\,.
\end{equation}
This is related to the Riemann curvature tensor calculated using the connection coefficients $\Gamma^\lambda{}_{\mu \nu} $ which are in turn related to the spin connection  through the tetrad postulate. In particular, the Ricci scalar is $R(\Gamma) = e_{a}^{\mu} e_{b}^{\nu} \tensor{F}{_{\mu\nu}}{^{ab}}(A)\,,$ where $e_{a}^{\mu}$ satisfy $e_{a}^{\mu}e^{b}_{\mu}=\delta^b_a.$ Thus in the absence of matter, the action for gravity is 
\begin{equation} 
	{S_{\textit{ VEP}}=\frac{1}{2\kappa} \int d^{4} x\,|e|\, e_{a}^{\mu} e_{b}^{\nu} \tensor{F}{_{\mu\nu}}{^{ab}}\,,\qquad }
\end{equation}
{where $\kappa = 8\pi G\,.$} Varying the action with respect to the connection $\tensor{A}{_\mu}{^{ab}}$, we find after some calculation that
$A_\mu{}^{ab} = \omega_\mu{}^{ab}\,,$ the spin connection for which the  $\Gamma^\lambda{}_{\mu \nu} $ are the torsion-free Christoffel symbols.
$\omega_\mu{}^{ab}$ are fully expressible in terms of the $e^\mu_a,\, e^a_\mu$ and their derivatives. 
Varying the $e^\mu_a$ produces the tetrad version of Einstein equations in vacuum. If there are minimally coupled bosonic fields in the theory, the equation for $A_\mu{}^{ab}$ remains the same, leading to the same solution.

The situation changes when fermions are included. Then the action becomes
%
%
\begin{equation}\label{action.1}
	S = \int |e| d^4x \left(\frac{1}{2\kappa} F_{\mu\nu}{}^{ab}(A) e^\mu_a e^\nu_b +  \frac{1}{2} \left( i\bar{\psi} \slashed{D}\psi + h. c.\right) - m\bar{\psi}\psi \right) \,.
\end{equation}
Now the solution for the spin connection does not produce a symmetric $\Gamma\,.$ If we split  the spin connection as $A_\mu{}^{ab} = \omega_\mu{}^{ab} +\Lambda_\mu^{ab}\,$ and vary $\Lambda$ independently of the $e^\mu_a\,,$
the derivatives of $\Lambda$ drop out of the action and $\Lambda$ satisfies an algebraic equation, 
\begin{equation}\label{spin-connection}
	\Lambda^{ab}_\mu = -\frac{\kappa}{4}\tensor{\epsilon}{^{ab}}{_{cd}}e^c_{\mu}\,\bar{\psi}\gamma^d\gamma^5\psi \,.
\end{equation}
This field corresponds to an antisymmetric part in $\Gamma\,$ and would be normally identified with spacetime torsion. However, since there is no derivative of $\Lambda$ in the action, and because $\Lambda$ can be written in terms of other fields without derivatives, this expression can be safely put back into the action. Then $\Lambda$ is eliminated from the action, leaving a 4-fermion interaction coupled to usual torsion-free gravity. The Dirac equation also becomes nonlinear,~\cite{Finkelstein:1960,Gasperini:2013,Gursey:1957,Hehl:1971qi}
\begin{equation}\label{DE1}
	i\slashed{\partial} \psi + \frac{1}{4} \omega_\mu{}^{ab}\gamma^\mu \sigma_{ab}\psi
	- m\psi 		- \frac{3\kappa}{8}\left(\bar{\psi}\gamma^a\gamma^5\psi\right)\gamma_a\gamma^5\psi = 0\,.
\end{equation}
We note that while $\omega_\mu^{ab}$ vanishes in flat spacetime, the nonlinear term does not. This is not a paradox, since spacetime is only approximately flat when matter is present. For laboratory experiments, i.e. in all cases where we assume a flat spacetime, $\omega_\mu^{ab}$ vanishes only approximately.

So far, we considered only one species of fermions. However, all species of fermions should be included in the action of Eq.~(\ref{action.1}). Furthermore, different species of fermions can couple independently to the field $\Lambda_\mu^{ab}$ -- with different coupling constants -- there is no symmetry which prevents that. Even more interestingly, $\Lambda$ can couple even to massless fermions -- for example, a purely left-handed neutrino. This implies that $\Lambda$ will generically couple to left- and right-handed components of each fermion with different coupling strengths. Thus we can write the action of fermions in curved spacetime as~\cite{Chakrabarty:2019cau}
\begin{align}
	{\mathscr L}_\psi &= \sum\limits_{i}	\left[\frac{i}{2}\bar{\psi}_i\gamma^\mu\partial_\mu\psi_i - 
	\frac{i}{2}\partial_\mu\bar{\psi}_i\gamma^\mu\psi_i 	- \frac{1}{4} \epsilon_{abcd}\omega_{\mu}{}^{ab} e^{\mu c}  \,
	\bar{\psi}_i \gamma^d\gamma^5 \psi_i\, - m\bar\psi_i\psi_i\, \right. \notag \\ 
	&\qquad \qquad \left.
	- \frac{1}{4}\epsilon_{abcd} \Lambda_{\mu}{}^{ab} e^{\mu c}
	\left(\tilde{\lambda}_{iL}\bar{\psi}_{iL} \gamma^d \psi_{iL} + \tilde{\lambda}_{iR}\bar{\psi}_{iR} \gamma^d\psi_{iR}\right)
	\right]\,,\label{L_psi_all}
\end{align}
where the sum runs over all species of fermions. The curvature term in the action~(\ref{action.1}) remains as before, so we find 
\begin{equation}\label{chiral.torsion}
	\Lambda_{\mu}{}^{ab} = -\frac{\kappa}{4}\epsilon^{abcd}e_{c\mu} \sum\limits_i \left(\tilde{\lambda}_{iL}\bar{\psi}_{iL}\gamma_d \psi_{iL} + \tilde{\lambda}_{iR}\bar{\psi}_{iR}\gamma_d \psi_{iR}\right)\,.
\end{equation}
As before, we can put this solution back into the action to get a four-fermion interaction term,
\begin{equation}\label{4fermi}
{\mathscr L}_{4\psi} =	-\frac{3\kappa}{16}\left(\sum\limits_i \left(\tilde{\lambda}_{iL}\bar{\psi}_{iL} \gamma_a \psi_{iL} + \tilde{\lambda}_{iR}\bar{\psi}_{iR} \gamma_a \psi_{iR}\right)\right)^2\,.
\end{equation}
The spin connection is now purely $\omega_\mu^{ab}\,,$ the action no longer contains $\Lambda\,.$ Thus we have usual torsion-free gravity, but with a four-fermion interaction induced by the geometry. We will refer to this term and the associated coupling constants as \textit{geometrical}. 

This interaction term does not involve the background metric or $\omega_\mu^{ab}\,,$ so it will be present in usual laboratory experiments, where the spacetime is very close to flat. Of course, spacetime is not exactly flat in the presence of matter,
but for the kind of matter densities which appear in laboratory experiments, the effect on the metric is negligible. 
The geometrical interaction  will also contribute to the energy-momentum tensor and thus affect the spacetime through Einstein equations, again with a negligible effect on the metric. 
Thus what we find is that all fermions in ordinary experiments obey the Dirac equation in flat space, but including a cubic term from the four-fermion interaction.
{ Note that this interaction is not an extension of general relativity, nor does it come from some generalization of gravity. 
It arises because there is a local Lorentz symmetry at each point of spacetime -- gauging the Poincar\'e symmetry produces both gravity 
and torsion -- what we have is the natural outcome of trying to combine general relativity with fermions. On the other hand, 
the geometrical interaction itself is separate from gravitation, the curvature of spacetime does not directly affect 
it. Furthermore, the torsion generated by the fermions is 
totally antisymmetric and therefore does not affect geodesics. All point particles fall at the same rate in vacuum in a gravitational field 
and the equivalence principle is not violated.

A four-fermion interaction is known to be nonrenormalizable when considered as a quantum field theory in flat spacetime. Although the curved nature of spacetime is fundamental to the interaction that we have found, we are interested in its effect in laboratory experiments, where the spacetime is nearly flat. So the issue of renormalizability may be relevant at high energies. We will therefore restrict ourselves to low energy experiments where we can ignore loop effects.


} 

It is easy to see that parity is violated by this interaction. Let us write the four-fermion interaction in terms of vector and axial currents,
\begin{equation}\label{V-A}
	{\mathscr L}_{4\psi} =-\frac{3\kappa}{16}\Big[\sum_i \left(\tilde{\lambda}_i^V \bar{\psi}_i \gamma^a \psi_i + \tilde{\lambda}_i^A \bar{\psi}_i \gamma^a\gamma^5 \psi_i\right)\Big]^2\,, 
\end{equation}
where we have  written $\tilde{\lambda}_i^{V, A} = \frac{1}{2}(\tilde{\lambda}_{iR} \pm \tilde{\lambda}_{iL})$\,. This term clearly violates parity.
The form of the geometrical interaction shows that it cannot contribute to charged current processes, but may contribute to neutral current processes. In this paper, we consider two such processes where parity violation is observed -- namely, electron-electron scattering and electron-nucleon scattering. It should be possible to get an order-of-magnitude estimate on the upper bound for the geometrical coupling constants $\lambda\,.$

%
%
%
%

Let us explain why we expect an observable effect at accessible energies even though the interaction arises from spacetime geometry. 
We are not considering the quantization of spacetime --  this interaction  is independent of the tetrad fields or the metric. Also, the contorsion field
$\Lambda_\mu^{ab}$ is non-propagating, so it has no intrinsic scale and thus can be eliminated without introducing a scale into the theory.
So even if a quantum theory of gravity is found some day and the scale of that is $M_P$ as expected, the scale of the 
geometrical four-fermion interaction need not be the same. 
Moreover, since the torsion field $\Lambda_\mu^{ab}$\, is not a quantum field, the dimensionless coupling constants $\tilde{\lambda}_i$ 
with the fermions are not restricted to be small. However, the interaction term ${\mathscr L}_{4\psi}$ needs to be small compared to
the free Hamiltonian if we are to use it in perturbation theory.  Let us rewrite (\ref{V-A})
as 
\begin{equation}\label{V-A.new}
	{\mathscr L}_{4\psi} = -\frac{1}{2}\left[\sum_i \left({\lambda}_i^V \bar{\psi}_i \gamma^a \psi_i + {\lambda}_i^A \bar{\psi}_i 
	\gamma^a\gamma^5 \psi_i\right)\right]^2\,, 
\end{equation}
where we have rescaled the coupling constants as $\lambda_i = \sqrt{\frac{3\kappa}{8}}\tilde{\lambda}_i\,.$ Although $\sqrt{\kappa}$ 
appears in the numerator here, it does not mean that $\lambda_i \sim 10^{-19}$ GeV$^{-1}$, because $\tilde{\lambda}_i$ can be 
arbitrarily large in principle. 
All we can say is that the couplings $\lambda_{i}$ cannot be determined 
by purely theoretical arguments, but should be fixed by comparing theoretical predictions with experimental observations. At low energies, 
a term exactly like (\ref{V-A.new}) appears as the limit of weak interactions. It is thus safe to assume that if the processes mediated by the
geometric interaction are not larger -- at least not much larger -- than weak interaction processes at low energies, we can calculate its 
effects by perturbative field theory and compare with experimental results.

\section{Electron-Electron Scattering}\label{Moller}
\begin{figure}[t]\includegraphics[width=3cm, height=3.6cm]{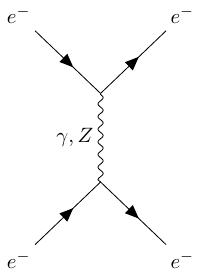} \hspace{1.5cm}\includegraphics[width=3cm, height=3.6cm]{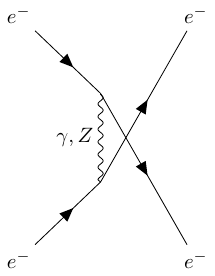}\hspace{1.5cm}\includegraphics[width=3cm, height=3cm]{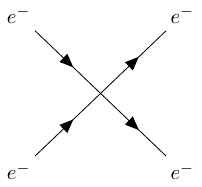}
	\caption{Tree level diagrams for M{\o}ller scattering}
	\label{Moller.fig}
\end{figure}

%
As explained above, we can work with flat space quantum field theory even for the geometrical interaction. At tree level, we have the diagrams of Fig.~\ref{Moller.fig}  due to photon exchange, $Z$ exchange, and geometrical interactions for M{\o}ller scattering $e^-(p_1)+e^-(p_2)\rightarrow e^-(p_3)+e^-(p_4)$.
The QED contribution to the amplitude is $\mathscr{M}^{e-e}_\gamma=\mathscr{M}_1-\mathscr{M}_2$, where
\begin{subequations}
\begin{equation}\mathscr{M}_1=\frac{e^2}{(p_1-p_3)^2}[\bar{u}_3\gamma^{\mu}u_1][\bar{u}_4 \gamma_{\mu}u_2]\,,\end{equation}
\begin{equation}\mathscr{M}_2=\frac{e^2}{(p_1-p_4)^2}[\bar{u}_4\gamma^{\mu}u_1][\bar{u}_3\gamma_{\mu}u_2 ]\,.\end{equation}
\end{subequations}
The subscripts on the spinors indicate spin and momentum labels of the electrons. 
Similarly, writing the $Z$ vertex of the electron as $-\frac{ie}{\sin 2\theta_W}\gamma^\mu(c^V_e-c^A_e\gamma^5)\,,$ the weak interaction contribution to the amplitude can be written as $\mathscr{M}^{e-e}_Z=\mathscr{M}_3-\mathscr{M}_4$, where 
\begin{subequations}
\begin{equation}
	\mathscr{M}_3= \frac{e^2}{\sin^22\theta_W ((p_1-p_3)^2 - M_Z^2)}\left[\bar{u}_3 \gamma^{\mu}(c^V_e-c^A_e\gamma^5)u_1\right]\left[\bar{u}_4\gamma_{\mu}(c^V_e-c^A_e\gamma^5)u_2\right]\,,
\end{equation}
\begin{equation}
	\mathscr{M}_4= \frac{e^2}{\sin^22\theta_W ((p_1-p_4)^2 - M_Z^2)}\left[\bar{u}_4 \gamma^{\mu}(c^V_e-c^A_e\gamma^5)u_1\right]\left[\bar{u}_3\gamma_{\mu}(c^V_e-c^A_e\gamma^5)u_2\right]\,,
\end{equation} 
\end{subequations}
where $c^V_e=-\frac{1}{2}+2\sin^2\theta_W$  and $c^A_e=-\frac{1}{2}$\,.
To these, we add the contribution due to the geometrical 4-fermion interaction, 
which we write as $\mathscr{M}^{e-e}_{\tau}=\mathscr{M}_5-\mathscr{M}_6\,,$ with
\begin{subequations}
	\begin{align}
		\mathscr{M}_5  &=[\Bar{u}_3\gamma^\mu({\lambda}^V_e+{\lambda}^A_e\gamma^5)u_1][\Bar{u}_4\gamma_\mu({\lambda}^V_e+{\lambda}^A_e\gamma^5)u_2]\,,\\
%
	\mathscr{M}_6 &=[\Bar{u}_4\gamma^\mu({\lambda}^V_e+{\lambda}^A_e\gamma^5)u_1][\Bar{u}_3\gamma_\mu({\lambda}^V_e+{\lambda}^A_e\gamma^5)u_2]\,.
	\end{align}
\end{subequations}
The total amplitude for $e^-$--$e^-$ elastic scattering is $\mathscr{M}=\mathscr{M}^{e-e}_\gamma+\mathscr{M}^{e-e}_Z+\mathscr{M}^{e-e}_{\tau}$\,.


Although the QED contribution dominates at low energies, the interference with neutral current and the torsional 4-fermion interaction should show a signature of parity violation. We consider M{\o}ller scattering of polarized electrons where left or right-handed electrons scatter from an unpolarized target. The measure of parity violation is
\begin{equation}
	A_{PV}=\frac{d\sigma_R-d\sigma_L}{d\sigma_R+d\sigma_L}\,,
\end{equation}
where $d\sigma_{R(L)}$ is the differential cross section for right-handed (left-handed) electrons scattering off an unpolarized electron target~\cite{Derman:1979zc}. 

We take electron $1$ to be polarized and electron 2 to be the unpolarized target and the cross sections are calculated for $\sqrt{s}\gg m_e$\,. 
The search for parity violation in M{\o}ller scattering is usually done at energies well below the $Z$ mass. We can therefore  replace  $(q^2 - M_Z^2)$ by $(-M_Z^2)\,$ in the neutral current amplitudes. The dominant contribution to the numerator of $A_{PV}$ comes from the interference of the $\gamma$ exchange diagram with the neutral current and the geometrical four-fermion interaction diagrams. This contribution is calculated to be
\begin{equation}
	\overline{\left|{\mathscr{M}}_R\right|^2}-\overline{\left|{\mathscr{M}}_L\right|^2}\approx {512\pi\alpha}(\sqrt{2}G_F c^V_ec^A_e+ \lambda^V_e\lambda^A_e)(p_1\cdot p_2)(p_3\cdot p_4)\left[\frac{1}{(p_1-p_3)^2}+\frac{1}{(p_1-p_4)^2}\right]\,,
\end{equation}
%
where the bar on top of ${\left|{\mathscr{M}}_{R,L}\right|^2} $ denotes averaging over the initial spins and sum over final spins. 
The phase space factors cancel, so for the differential cross sections $d\sigma_{R,L}$, it is sufficient to calculate the spin sums $\overline{\left|{\mathscr{M}}\right|^2}_{R,L} = \frac{1}{2}\sum|\mathscr{M}|_{R,L}^2$ where the sum is over the spins of all
electrons, but the polarization ${R,L}$ of electron 1 is picked by inserting $\frac{1}{2}(1\pm\gamma_5)\,.$ 

In order to compare with the results of~\cite{Derman:1979zc}, we work in the center-of-momentum (CM) frame and define the variable {$y := -\frac{q^2}{s} \equiv -\frac{(p_1-p_3)^2}{s}=\sin^2\frac{\theta}{2}$}, where $\theta$ is the scattering angle in the CM frame. In terms of $y$ and $s$ we have the following relations:
\begin{align}
	(p_1-p_4)^2 &= s(y-1)\,, \notag\\
	p_1\cdot p_2 = p_3\cdot p_4 &=\frac{1}{2}\,s\,, \notag\\
	p_1\cdot p_4 =p_2\cdot p_3 &=\frac{1}{2}\,s(1-y)\,, \notag \\
	p_1\cdot p_3 = p_2\cdot p_4 &=\frac{1}{2}\,sy\,. 
\end{align}
The numerator of $A_{PV}$ is therefore
\begin{equation}
\overline{\left|{\mathscr{M}}_R\right|^2} - \overline{\left|{\mathscr{M}}_L\right|^2} \approx-{128s\pi\alpha}\frac{(\sqrt{2}G_F c^V_ec^A_e+ \lambda^V_e\lambda^A_e)}{ y(1-y)}\,.
\end{equation}
For the denominator of $A_{PV}$ we need only the QED contribution, as the other contributions are much smaller, so
\begin{equation}
\overline{\left|{\mathscr{M}}_R\right|^2} + \overline{\left|{\mathscr{M}}_L\right|^2}	\approx
\frac{64\pi^2\alpha^2(1+y^4+(1-y)^4)}{y^2(1-y)^2}\,.
%
\end{equation}
%
%
%
%

Thus the parity-violating asymmetry for electron-electron scattering is
\begin{equation}\label{asym.1}
A_{e-e}=-\frac{2s}{\pi\alpha}(\sqrt{2}G_F c^V_ec^A_e+ \lambda^V_e\lambda^A_e)\frac{y(1-y)}{1+y^4+(1-y)^4}\,.
\end{equation}
%

\section{Electron-Deuteron Scattering}\label{e-D}

Parity violating deep inelastic lepton-hadron scattering is a useful tool for probing the structure of the nucleon. Earlier it was used to distinguish between the standard model and other candidates. For our purposes, we can use them to establish bounds on the quark couplings. In this section, we take up electron-deuteron scattering and calculate the asymmetry $A_{PV}$, following  the treatment of~\cite{Cahn:1977uu}.

To write down an expression for the parity-violating asymmetry for electron-deuteron interaction, we start with the amplitudes for $e^-(p_1)+q_i(p_2)\rightarrow e^-(p_3)+q_i(p_4)\,,$ an electron scattering off a quark $q_i$  where $ i=u,d\,.$
\begin{figure}[ht]\includegraphics[width=3cm, height=3cm]{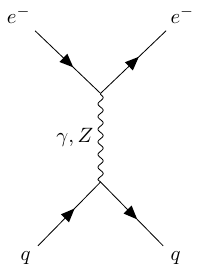} \hspace{3cm}\includegraphics[width=3cm, height=3cm]{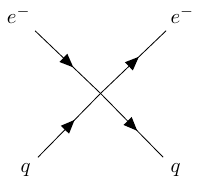}
	\caption{Tree level diagrams for electron scattering off a quark}
	\label{diag-e-q}
\end{figure}

There are three diagrams at the tree level, shown in Fig.~\ref{diag-e-q}, with amplitudes
\begin{subequations}
\begin{align}
\mathscr{M}_{\gamma}^{e-q} &=-\frac{e^2Q_i}{q^2}\left[\bar{u}_3\gamma^{\mu}u_1][\bar{u}_4 \gamma_{\mu}u_2\right]\,,  \\
%
	\mathscr{M}_Z^{e-q} &=\frac{e^2}{\sin^22\theta_W (q^2 - M_Z^2)}\left[\bar{u}_3 \gamma^{\mu}(c^V_e-c^A_e\gamma^5)u_1][\bar{u}_4\gamma_{\mu}(c^V_i-c^A_i\gamma^5)u_2\right]\,,\\
%
	\mathscr{M}_{\tau}^{e-q} &=\left[\bar{u}_3 \gamma^{\mu}(\lambda^V_e+\lambda^A_e\gamma^5)u_1][\bar{u}_4\gamma_{\mu}(\lambda^V_i+\lambda^A_i\gamma^5)u_2\right]\,,
\end{align} 	
\end{subequations}
where $q^2=(p_1-p_3)^2$, $Q_u=+2/3\,, Q_d=-1/3\,,$ the electron couplings $c^V_e$ and $c^A_e$ are as before, while the quark coupling are $c^V_u = \frac{1}{2} - \frac{4}{3}\sin^2\theta_W$ and $c^V_d = -\frac{1}{2} + \frac{2}{3}\sin^2\theta_W$ for the vector couplings, and $c^A_u = \frac{1}{2}\,,\, c^A_d = -\frac{1}{2}$ for the axial couplings.

At this stage, we can use the explicit helicity eigenstates to write out the amplitudes $\mathscr{M}^{\gamma}_{RR}$, $\mathscr{M}^{\gamma}_{RL}$, $\mathscr{M}^{\gamma}_{LR}$, $\mathscr{M}^{\gamma}_{LL}$ and also for the neutral current and torsional interaction where the first and second subscripts denote the helicity of the incoming electron and quark respectively. 
Then in the CM frame and neglecting the masses of the particles, we have~\cite{Derman:1979zc}
\begin{eqnarray}
	&\bar{u}_{\lambda^\prime}(\mathbf{p}_3)\gamma^{\mu}u_{\lambda}(\mathbf{p}_1)=\sqrt{s}\left[\cos\left(\frac{\theta}{2}\large\right),\, \sin\left(\frac{\theta}{2}\large\right),\,-(-1)^{(1+\lambda)/2}i\sin\left(\frac{\theta}{2}\large\right),\,\cos\left(\frac{\theta}{2}\large\right)\right]\delta_{\lambda\lambda^\prime}\,,\\
	&\bar{u}_{\lambda^\prime}(\mathbf{p}_3)\gamma^{\mu}\gamma^5u_{\lambda}(\mathbf{p}_1)=-(-1)^{(1+\lambda)/2}\bar{u}_{\lambda^\prime}(\mathbf{p}_3)\gamma^{\mu}u_{\lambda}(\mathbf{p}_1)\,,\\
	&\bar{u}_{\lambda^\prime}(\mathbf{p}_4)\gamma^{\mu}u_{\lambda}(\mathbf{p}_2)=\sqrt{s}\left[\cos\left(\frac{\theta}{2}\large\right),\, -\sin\left(\frac{\theta}{2}\large\right),\,-(-1)^{(1+\lambda)/2}i\sin\left(\frac{\theta}{2}\large\right),\,-\cos\left(\frac{\theta}{2}\large\right)\right]\delta_{\lambda\lambda^\prime}\,,\\
	&\bar{u}_{\lambda^\prime}(\mathbf{p}_4)\gamma^{\mu}\gamma^5u_{\lambda}(\mathbf{p}_2)=-(-1)^{(1+\lambda)/2}\bar{u}_{\lambda^\prime}(\mathbf{p}_4)\gamma^{\mu}u_{\lambda}(\mathbf{p}_2)\,,
\end{eqnarray}
where $\theta$ is the scattering angle, $\lambda=+1$ for right-handed particles and $\lambda=-1$ for left-handed particles.

The polarized cross-sections are given by $d\sigma_{\lambda\lambda^\prime}\sim \abs{\mathscr{M}^{\gamma}_{\lambda\lambda^\prime}+\mathscr{M}^{Z}_{\lambda\lambda^\prime}+\mathscr{M}^{\tau}_{\lambda\lambda^\prime}}^2$ (up to phase space factors). We find
\begin{subequations}
\begin{align}
d\sigma_{RR}&\sim\left(\frac{Q^\gamma_eQ^\gamma_i}{q^2}+\frac{Q^Z_{Re}Q^Z_{Ri}}{q^2-M^2_Z}+Q^\tau_{Re}Q^\tau_{Ri}\right)^2\\ d\sigma_{RL}&\sim\left(\frac{Q^\gamma_eQ^\gamma_i}{q^2}+\frac{Q^Z_{Re}Q^Z_{Li}}{q^2-M^2_Z}+Q^\tau_{Re}Q^\tau_{Li}\right)^2(1-y)^2\\
d\sigma_{LR}&\sim\left(\frac{Q^\gamma_eQ^\gamma_i}{q^2}+\frac{Q^Z_{Le}Q^Z_{Ri}}{q^2-M^2_Z}+Q^\tau_{Le}Q^\tau_{Ri}\right)^2(1-y)^2\\
d\sigma_{LL}&\sim\left(\frac{Q^\gamma_eQ^\gamma_i}{q^2}+\frac{Q^Z_{Le}Q^Z_{Li}}{q^2-M^2_Z}+Q^\tau_{Le}Q^\tau_{Li}\right)^2\,,
\end{align}	
\end{subequations}
where $y = \sin^2\frac{\theta}{2}\,.$
The geometrical contribution does not contain a propagator, being a contact interaction. 
Here the $Q^\gamma$ and $Q^Z$ are the electromagnetic and weak charges of the particles, defined as 
\begin{align}
	Q^\gamma_e &=-e\,, 
	& Q^\gamma_i =eQ_i\,, \qquad\qquad\qquad \notag \\
	Q^Z_{Re} &=\frac{e}{\sin2\theta_W}(c^V_e-c^A_e)\,, 
& 	Q^Z_{Le} =\frac{e}{\sin2\theta_W}(c^V_e+c^A_e)\,, \notag \\
	Q^Z_{Ri} &=\frac{e}{\sin2\theta_W}(c^V_i-c_i^A)\,, 
	&Q^Z_{Li} =\frac{e}{\sin2\theta_W}(c^V_i+c_i^A)\,, 
\end{align}
and the $\tau$ charges are defined as
\begin{align}
	Q^\tau_{Re} &=(\lambda^V_e+\lambda^A_e)\,, 
\qquad 	&Q^\tau_{Le} =(\lambda^V_e-\lambda^A_e)\,,  \notag \\
	Q^\tau_{Ri} &=(\lambda^V_i+\lambda^A_i)\,, 
\qquad 	&Q^\tau_{Li} =(\lambda^V_i-\lambda^A_i) \,.
\end{align}
%
%

We wish to find an expression for $A_{PV}=\frac{d\sigma_R-d\sigma_L}{d\sigma_R+d\sigma_L}$ 
for scattering of polarized electrons from an unpolarized nucleon target like the deuteron. 
Then we have to take a sum over parton types as {$d\sigma_R=\frac{1}{2}\sum_{i}f_i(x)(d\sigma^{(i)}_{RR}+d\sigma^{(i)}_{RL})$ and $d\sigma_L=\frac{1}{2}\sum_if_i(x)(d\sigma^{(i)}_{LR}+d\sigma^{(i)}_{LL})$} due to the presence of 
both up and down quarks inside the nucleus, where $x$ is the Bjorken scaling variable.
However, for a deuteron target with an equal 
number of up and down quarks, we may take $f_u(x)=f_d(x)$ and then the $x$ dependence drops out of $A_{PV}$. 
Keeping terms up to order $q^2/M_Z^2$, we arrive at the result 
\begin{equation}
	A_{e-D}=A^{EW}_{e-D}+A^{\tau}_{e-D}\,,
\end{equation}
where
\begin{equation}
A^{EW}_{e-D}=-\frac{q^2}{M^2_Z}\frac{\sum_iQ^\gamma_i[(Q^Z_{Re}Q^Z_{Ri}-Q^Z_{Le}Q^Z_{Li})+(Q^Z_{Re}Q^Z_{Li}-Q^Z_{Le}Q^Z_{Ri})(1-y)^2]}{Q^\gamma_e\sum_i(Q^\gamma_i)^2[1+(1-y)^2]}
\end{equation}
is the electroweak contribution~\cite{Cahn:1977uu}, and 
\begin{equation}
	A^{\tau}_{e-D}=\frac{q^2\sum_iQ^\gamma_i\left[(Q^\tau_{Re}Q^\tau_{Ri}-Q^\tau_{Le}Q^\tau_{Li})+(Q^\tau_{Re}Q^\tau_{Li}-Q^\tau_{Le}Q^\tau_{Ri})(1-y)^2\right]}{Q^\gamma_e\sum_i(Q^\gamma_i)^2[1+(1-y)^2]}\,.
\end{equation}
is the contribution from the geometrical or torsional interaction.

{At this stage, we make a simplifying assumption that the torsional interaction maximally violates parity, with the right-handed $\tau$-charges being zero, leaving only the left-handed ones. (Then for any fermion, $\lambda^V = - \lambda^A = Q^\tau/2\,.$) }
Moreover, we assume that the quark geometrical couplings are flavor-independent,  {$Q^\tau_u = Q^\tau_d$\,.} 
Then, writing out the charges in terms of the weak mixing angle, we have
\begin{equation}\label{asym.2}
	A_{e-D}=\frac{q^2}{20\pi\alpha\left[1+(1-y)^2\right]}\left[{G_F}\sqrt{2}\left({9 - \sin^2\theta_W\left[28 - 8(1 - y)^2\right]}\right)
	+{12\lambda_e\lambda_q}\right]\,.	
\end{equation} 
%
%
We calculated this assuming $-q^2\ll M^2_Z$\,, where $q$ is the (Euclidean) momentum transferred from the electron to the target.
For larger values of $-q^2$\,, corrections due to  electron-nucleon deep inelastic scattering must be taken into account. We refer the reader to~\cite{Hobbs:2008zc,Brady:2011zc,Belitsky:2011zc} for corrections due to finite $q^2$ as well as other corrections . 

\section{Discussion}
In this paper, we have argued that space-time torsion, generated by the dynamics of fermions, contributes to the 
parity-violating asymmetry in fermion scattering, even when the curvature of spacetime is negligible. The issue 
of parity nonconservation in curved spacetime has been studied in the literature in the context of nonminimal coupling of fermions 
with gravity, starting with the Holst action~\cite{Freidel:2005zc, Perez:2006zc}. In contrast, we have considered
minimal coupling, which gives rise to a contorsion field independent of usual gravity. We have chosen the coupling between 
fermions and contorsion  to be nonuniversal and also different for left- and right-handed fermions.

While this effect is small, it is independent of other known interactions including gravity, so there is no way of predicting the 
size of the effect by theoretical reasoning. We can get an order of magnitude estimate for the geometric coupling constants 
$\lambda$ by comparing the measured values of $A_{PV}$ with the results we have obtained.

For example, consider $A_{PV}$ for M{\o}ller scattering, measured by the SLAC E158 Collaboration~\cite{SLACE158:2003onx}. 
For this measurement, the average values of the kinematic variables were {$Q^2 := -q^2 = 0.026$ GeV$^2$ }and $y = Q^2/s\simeq 0.6$\,, where 
$s$ is the square of the $CM$ energy.
The result found was $A^{\text{ex}}_{e-e}=-175 \pm 30$(stat) $\pm 20$(sys) ppb. A later report~\cite{Anthony:2005zc} improved this to 
$A^{\text{ex}}_{e-e}=-131 \pm 14$(stat) $\pm 10$(sys) ppb.
From Eq.~(\ref{asym.1}), 
we see that the Standard Model contribution to the asymmetry at tree level is given by
\begin{equation}\label{asym.1.SM}
	A^{\text{SM}}_{e-e} = -\frac{G_F Q^2}{\sqrt{2}\pi\alpha}(1-4\sin^2\theta_W)\frac{(1-y)}{1+y^4+(1-y)^4}\,.
\end{equation}
%
%

{ Using the values of $G_F$ and $\alpha$ at low energy~\cite{Workman:2022zc}, and also  the 
standard model expectation of $\sin^2\theta_W(Q)= 0.2381$ at $Q^2 = 0.026$ GeV$^2$~\cite{Anthony:2005zc}, 
we arrive at the standard model value for the asymmetry at tree level $A^{\text{SM}}_{e-e,\text{tree}}=-154.17$ ppb. One loop electroweak radiative corrections to the M{\o}ller asymmetry was investigated in~\cite{Czarnecki:1995zc}. Using the formula given there, we find that for the present kinematics and with $m_H=125$ GeV, the asymmetry at one loop is $A^{\text{SM}}_{e-e,\,\text{1-loop}}=-146.06$ ppb.}

{A crude estimate for the size of the  geometrical coupling constants may be obtained by comparing 
	the total theoretical prediction with the  experimental value, which gives 
%
\begin{equation}
	\abs{-\frac{2Q^2}{\pi\alpha}\lambda^V_e\lambda^A_e\frac{(1-y)}{1+y^4+(1-y)^4}}<\abs{A^{\text{ex}}_{e-e}-A^{\text{SM}}_{e-e,\,\text{1-loop}}}=15.06\times 10^{-9},
\end{equation}
%
From this we estimate
\[
\abs{\lambda^V_e\lambda^A_e}< 19.17\times 10^{-9}{\rm GeV}^{-2}\,,
\] 
%
which is equivalent to
\[
\abs{\frac{\lambda^V_e\lambda^A_e}{\sqrt{2}G_Fc^V_ec^A_e}}< 0.098\,.
\]
%
($\lambda^A_e = -\lambda^V_e$ if the geometrical interaction maximally violates parity and involves only left chiral electrons. ) We note that the M{\o}ller asymmetry at two loops has been studied in~\cite{Du:2021zc} including closed fermion loops. These numbers will be tested in precision measurements in the upcoming MOLLER experiment~\cite{Mammei:2012,Benesch:2014fb}.}
%

%
{For the purpose of comparing our other calculation, we note that the parity asymmetries for an electron scattering off a deuteron were measured by the Jefferson Lab PVDIS Collaboration in~\cite{Wang:2013zc,Wang:2014zc,Wang:2015zc}. The asymmetry reported at the average values of $Q^2=1.085$ GeV$^2$, $E=6.067$ GeV, {$E'= 3.66$ GeV,}  was $A^{\text{ex}}_{e-D}=-91.1 \pm 3.1$(stat) $\pm 3.0$(sys) ppm. Here, $E$ is the incident electron energy and {$E^{\prime}$ is the scattered electron energy}. From Eq.~(\ref{asym.2}), we have the standard model contribution to the asymmetry
\begin{equation}
   A^{\text{SM}}_{e-D}=-\frac{G_FQ^2}{10\sqrt{2}\pi\alpha}\frac{[9 - \sin^2\theta_W(28 - 8(1 - y)^2)]}{[1+(1-y)^2]}\,.
\end{equation} 
This can be calculated using the numerical values of the various couplings and parameters given in~\cite{Wang:2014zc,Wang:2015zc}. 
The fine structure constant $\alpha$ is evolved to the measured 
$Q^2$ value from $\alpha(Q^2=0)= 1/137.036$, taking into account purely electromagnetic vacuum polarization, to get $\alpha=1/134.45$~\cite{Wang:2014zc,Wang:2015zc}. The value of the Fermi constant $G_F$ used is the same as that at low energy.  
Using { $y=(E-E^{\prime})/E\simeq 0.397$\,, we calculate}
the standard model value for the asymmetry $A^{\text{SM}}_{e-D}= {-88.57}$ ppm. The asymmetry with finite $Q^2$ corrections is given 
in~\cite{Hobbs:2008zc,Wang:2014zc,Wang:2015zc}, the corrected standard model value corresponding to these corrections being
$A^{\text{SM}}_{e-D}=-88.18$ ppm for the kinematics above. Here we have worked with the valence quark approximation by neglecting 
the effects of sea quarks completely. Thus we can make an estimate of the geometrical couplings as 
\begin{equation}
	\abs{-\frac{12Q^2\lambda^e\lambda^q}{20\pi\alpha[1+(1-y)^2]}}<\abs{A^{\text{ex}}_{e-D}-A^{\text{SM}}_{e-D}}=2.92 \times 10^{-6}\,,
\end{equation}
%
which is equivalent to 
\[
\abs{\lambda^e\lambda^q}<{142.92} \times 10^{-9}\, {\rm GeV}^{-2}\,.
\] 
%
 }
 
Parity violation has also been observed in  proton-proton elastic scattering~\cite{Berdoz:2003zc}. Analogous to, and in contrast with, parity 
violation in electron scattering, these are a result of interference between the amplitudes of strong and weak interactions~\cite{Milsteina:2020knt, Milstein:2020lzv}. While the geometrical interaction should contribute to that as well, we do not consider it here as calculations for strong 
interactions are done under various approximations which are beyond the scope of this work.

\appendix

\end{document}